\documentclass{article}

    \PassOptionsToPackage{numbers,compress,sort}{natbib}

\usepackage[final]{neurips_2022_ml4ps}

\usepackage[utf8]{inputenc} 
\usepackage[T1]{fontenc}    
\usepackage{hyperref}       
\usepackage{url}            
\usepackage{booktabs}       
\usepackage{amsfonts}       
\usepackage{nicefrac}       
\usepackage{microtype}      
\usepackage{xcolor}         
\usepackage{amsmath}
\usepackage{graphicx}	

\title{HIGlow: Conditional Normalizing Flows for High-Fidelity HI Map Modeling}

\author{%
  Roy Friedman 
    \\
  Department of Computer Science\\
  The Hebrew University of Jerusalem\\
  Jerusalem, Israel \\
  \texttt{roy.friedman@mail.huji.ac.il} \\
   \And
   Sultan Hassan~\thanks{Department of Astrophysical Sciences, Princeton University, Princeton, NJ, 08544}\ ~\thanks{Department of Physics \& Astronomy, University of the Western Cape, Cape Town 7535}~\thanks{NHFP Hubble Fellow}\\
   Center for Computational Astrophysics\\
   Flatiron Institute\\
   New York, NY 10010 \\
   \texttt{shassan@flatironinstitute.org} \\
}

\begin{document}

\maketitle

\begin{abstract}
Extracting the maximum amount of cosmological and astrophysical information from upcoming large-scale surveys remains a challenge. This includes evaluating the exact likelihood, parameter inference and generating new diverse synthetic examples of the incoming high-dimensional data sets. 
In this work, we propose the use of normalizing flows as a generative model of the neutral hydrogen (HI) maps from the CAMELS project. Normalizing flows have been very successful at parameter inference and generating new, realistic examples. Our model utilizes the spatial structure of the HI maps in order to faithfully follow the statistics of the data, allowing for high-fidelity sample generation and efficient parameter inference.
\end{abstract}

\section{Introduction}


As the fields of cosmology and astrophysics enter the era of big-data, next-generation large-scale surveys, such as the Square Kilometer Array~\cite[SKA,][]{SKA}, the Hydrogen Epoch of Reionization Array~\cite[HERA,][]{HERA},  the Low Frequency Array~\cite[LOFAR,][]{LOFAR}, the Vera C. Rubin Observatory Legacy Survey of Space and Time~\citep[LSST,][]{LSST}, Nancy Grace Roman Space Telescope~\citep[Roman,][]{Roman}, Spectro-Photometer for the History of the Universe, Epoch of Reionization, and Ices Explorer~\citep[SPHEREx,][]{SPHEREx} and  Euclid~\citep{Euclid}, will enable imaging the neutral hydrogen (HI) in the early universe using unprecedented sensitivity and large fields of views. These surveys are expected to provide large amounts of high-dimensional data sets that require new tools, able to efficiently extract as much cosmological and astrophysical information as possible. A key challenge in the analysis of these new data sets lies in evaluating the exact likelihood of new samples in order to perform parameter inference that will enable translation of these growing observational efforts into tight theoretical constraints. Additionally, generating labeled synthetic large-scale HI maps, which faithfully capture the high-order statistics of the data, is equally important in order to facilitate accurate forecasting for upcoming surveys.

Traditional inference methods usually rely on the use of summary statistics, such as the spherically-averaged power spectrum, and are prone to loss of information due to data compression. However, a key lesson learnt from recent state-of-the-art convolutional neural networks (CNNs) is that performing inference at the field level is crucial to extracting the maximum amount of information~\citep{CNN1,CNN2}. 
While there exist several families of generative models, such as generative adversarial networks~\citep{GAN} (GANs) and variational autoencoders~\citep{VAE} (VAEs), they are not able to provide full access to the exact likelihood of high-dimensional data. On the other hand, normalizing flows~\citep{dinh2014nice, Glow, cGlow, SRFlow, papamakarios2021normalizing} are emerging as a powerful technique that provides access to high-dimensional likelihood and can generate high fidelity new samples. 

Recently, a normalizing flow model called HIFlow~\citep{HIFlow} that is based on conditional masked autoregressive flows was introduced as a generative model for simulated HI maps. While successful, HIFlow does not utilize the spatial structure of these HI maps and as such is unable to capture the full statistics of the data. In this work we propose a new model, called \emph{HIGlow}\footnote{Code available at \href{https://github.com/friedmanroy/HI-generation}{github.com/friedmanroy/HI-generation}}, which is a conditional version of Glow~\citep{Glow}, a powerful normalizing flow for modeling images. By using convolutional operations, our model is more suited to the spatial structure of the HI maps and is able to faithfully capture their high-frequency statistics. Additionally, using our conditional model we show how parameter inference is made possible using the posterior distribution of our model.

\subsection{Simulation}\label{sec:simulation}
We use simulations from the CAMELS project, recently introduced in~\citet{CAMELS}.  CAMELS is a suite of thousands of simulations run with state-of-the-art cosmological hydrodynamic galaxy formation models, by varying two cosmological parameters ($\Omega_m$ and $\sigma_8$) and four other parameters that modify the strength of stellar feedback (ASN1, ASN2) and black hole feedback (AAGN1, AAGN2) relative to the original IllustrisTNG and Simba simulations.
We focus our analysis on the Latin-hypercube (LH) set, which is a set of 1,000 simulations that explores this six-dimensional parameter space with uniform prior ranges.

\citet{HIFlow} showed that the distribution of the HI maps from the CAMELS simulations is largely dependent on the cosmological parameters $\Omega_m$ and $\sigma_8$. As such, we chose to focus on a model that is conditional on only these two parameters. Nonetheless, our model can be extended in order to use more conditional parameters, if needed.

\section{Methods}

Normalizing flows are a flexible framework for modelling probability distributions. Intuitively, normalizing flows utilize the change of variable formula in order to define a parametric transformation of a simple distribution into a more flexible distribution, in order to model the distribution of data. See~\citep{papamakarios2021normalizing} for an in-depth discussion of normalizing flows and their applications.


\begin{figure}
\begin{center}
    \includegraphics[width=.8\linewidth]{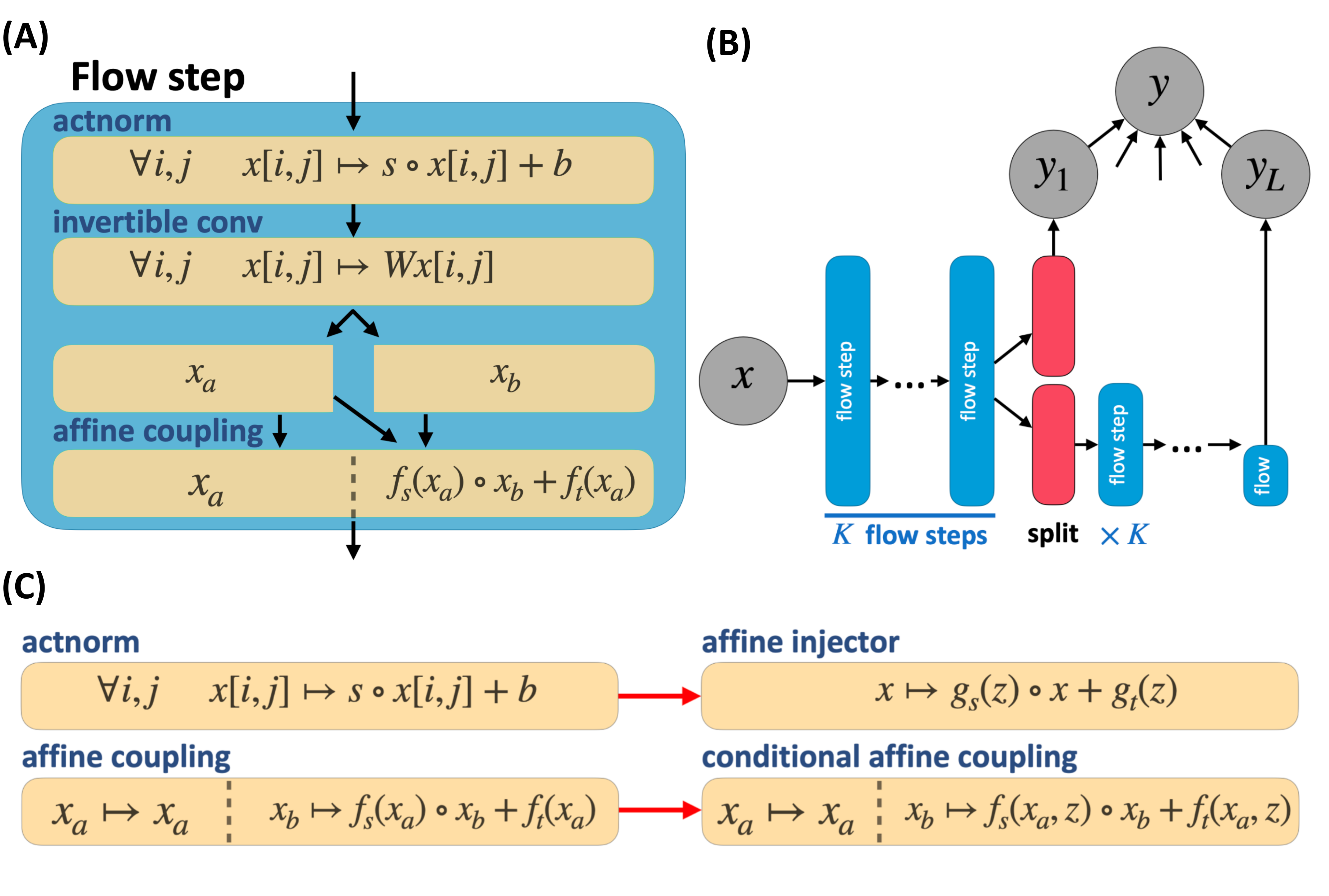}
\end{center}
  \caption{(A) normalizing flow modules used in Glow, (B) Glow architecture and (C) module changes in order to make Glow conditional}
\label{fig:flow-architecture}
\end{figure}

\subsection{Glow}\label{sec:glow}

Glow (\cite{Glow}) is a normalizing flow model that utilizes $1\times 1$ invertible convolutions in order to define invertible functions that take spatial information into account. The Glow architecture can be simplified into 3 modules which are stacked in order to create a single flow step, which are further stacked in order to get a flow block. These modules are:

{\bf Actnorm} which scales and translates the different channels in the input.

{\bf Invertible $1\times1$ convolutions} which act as a generalization of a permutation.

{\bf Affine coupling} which is the workhorse of the Glow architecture and allows for invertible affine transformations of the input.

The Glow architecture further utilizes the spatial structure of the images by squeezing the spatial dimension into the channel dimension (see~\citep{RealNVP}), which allows for more efficient splitting in the affine coupling layers and the use of a multi-scale approach; Figure~\ref{fig:flow-architecture}-A shows a schematic of the layers mentioned above while Figure~\ref{fig:flow-architecture}-B shows a schematic of Glow's architecture. The implementations of Glow's affine coupling layers utilize CNNs, making it particularly useful for modelling images.

\subsection{HIGlow}

In conditional modelling, we are given a dataset of pairs $\mathcal{D}=\left\{(z_i,x_i)\right\}_{i=1}^N$ where the $z_i$s are the underlying parameters for the generation of $x_i$. In this case, the task is to model the conditional distribution $p_x\left(x_i|z_i\right)$. The Glow model, as described above, does not have the capability to model the required conditional probabilities. 

In order to turn a normalizing flow model into a conditional model, it is enough to change the transformation into one that takes $z$, as well as the latent vector $y$, as an input: $x=f_\theta^{-1}(y,z)$.
In practice, this can be accomplished by changing the modules described in section~\ref{sec:glow}.
 
We follow the methods of \cite{SRFlow,cGlow} to turn Glow into a conditional model, using variants of the actnorm and affine coupling modules. Given the parameters $z$, these new layer types are:

\textbf{Affine injector}, a variant of the actnorm layer, with the following transformation:
\begin{equation}\label{eq:affine-injector}
    y=g_s(z)\circ x+g_t(z)
\end{equation}
where $g_s(\cdot)$ and $g_t(\cdot)$ are neural networks. This transformation was placed instead of the actnorm in Glow.

\textbf{Conditional affine coupling} is the same as the original affine coupling, only $z$ is added as an input to the affine transformation. This transformation was used instead of the standard affine couplings of Glow.

A schematic of both of the newly defined layers and how they compare to the standard Glow network can be seen in~\ref{fig:flow-architecture}-C.

In order to insert conditional information into the affine coupling layer, as described above, a new constant channel was added to the affine coupling's input for every parameter. Additionally, the affine injector in Equation~\ref{eq:affine-injector} is defined according to two functions: $g_s(z)$ and $g_t(z)$. In our implementation, both of these functions were implemented as multi-layer perceptrons (MLPs) with a single hidden layer from $z$ to a scalar.

\section{Results}

For all experiments, we used HIGlow with 6 flow blocks, each containing 20 flow steps on $64\times64$ pixel HI maps. Both the conditional affine coupling as well as the affine injectors had 1 hidden layer of width 16. The model was trained for 1000 epochs on one GPU.

\subsection{HI map generation}
\begin{figure*}
\begin{center}
    \includegraphics[width=\linewidth]{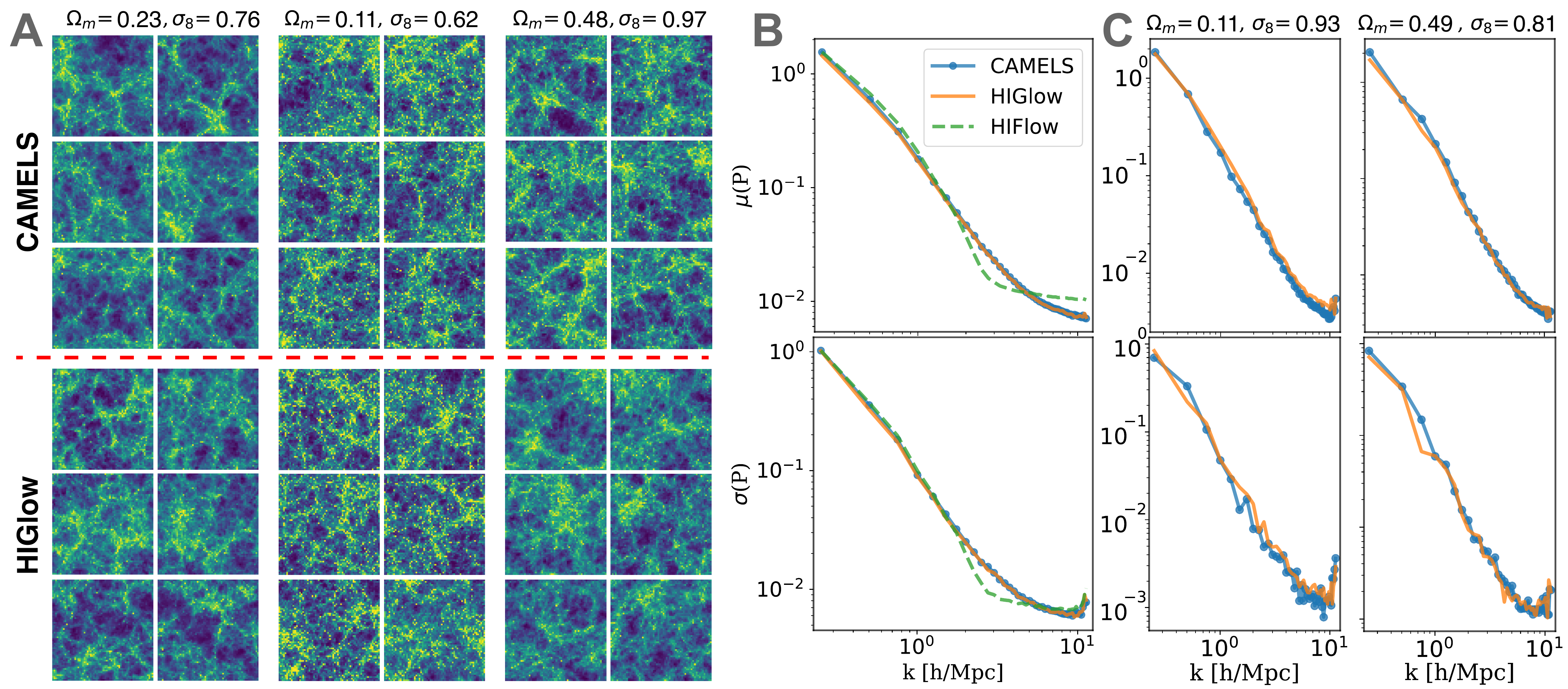}
\end{center}
  \caption{\textbf{(A)} Samples from CAMELS (top) next to samples generated from HIGlow (bottom) for different parameter settings. \textbf{(B)} Comparison of marginal power spectrum mean and standard deviation between CAMELS, HIGlow and HIFlow; \textbf{(C)} comparison of power spectra under different conditional parameters. }
\label{fig:generation}
\end{figure*}

HIGlow can be used in order to generate new HI maps, conditional on the cosmological parameters $\Omega_m$ and $\sigma_8$. As can be seen in Figure~\ref{fig:generation}-A, samples generated from HIGlow are visually very similar to those from the CAMELS simulation, even for different parameter settings.

To ensure that the statistics of the generated HI maps follow the statistics of the CAMELS simulation data, we compare the power spectra of both methods. Figure~\ref{fig:generation}-B shows the mean power spectrum and standard deviation from it as a function of spherical frequency of 500 maps unconditionally generated from HIGlow against 500 maps from CAMELS and 500 maps generated by HIFlow~\citep{CAMELS}, a previous approach using autoregressive flows in order to model HI maps from CAMELS. HIGlow is able to correctly model the high-frequency information in the HI maps, unlike HIFlow. In Figure~\ref{fig:generation}-C, the same curves can be seen for specific parameters values. In these HIGlow is able, again, to generate images with similar mean spectrum and standard deviation.

\subsection{Posterior distribution and parameter inference}
\begin{figure}
\begin{center}
    \includegraphics[width=\linewidth]{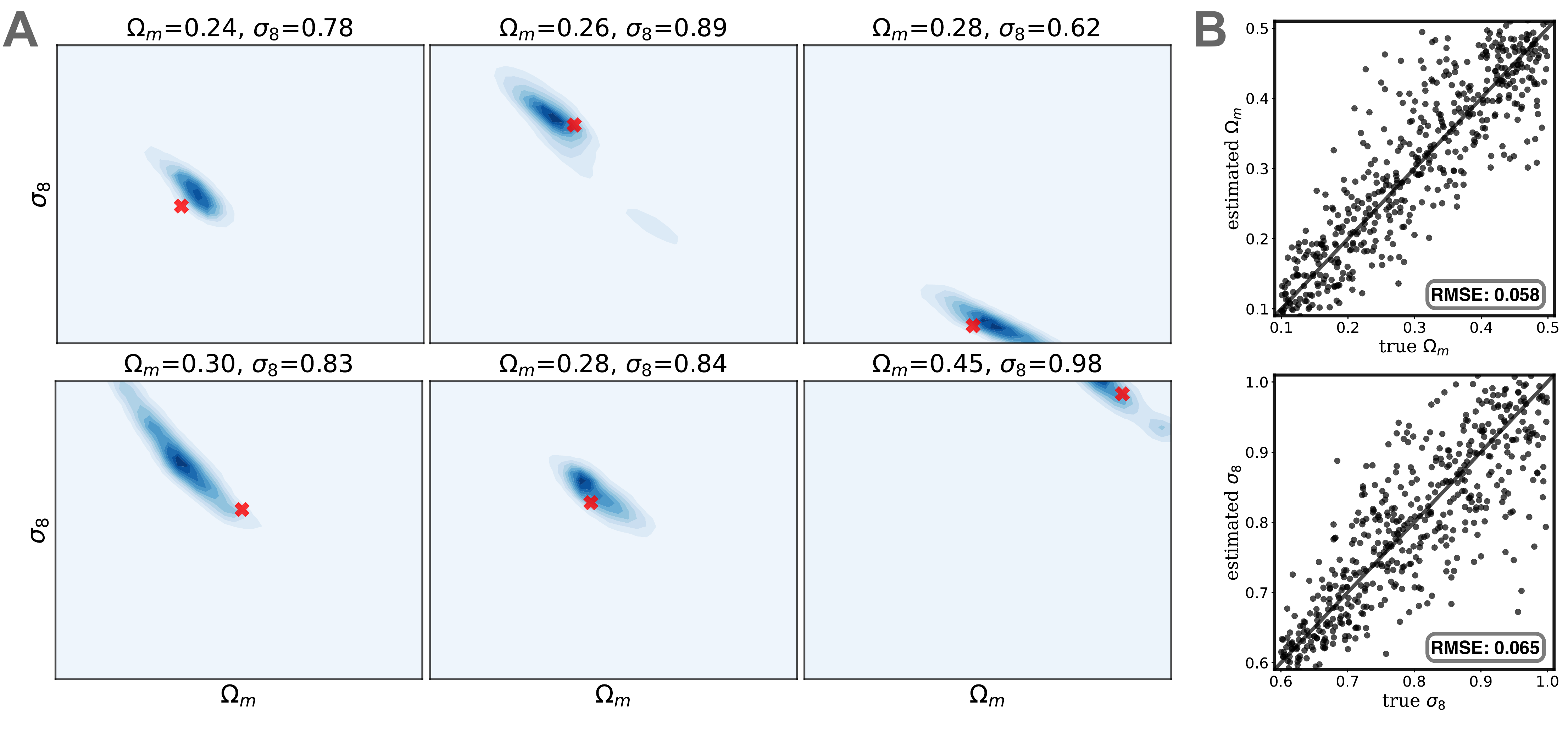}
\end{center}
  \caption{\textbf{(A)} Posterior distribution of HI maps with different parameters. \textbf{(B)} Estimated parameters using the mean of the posterior versus the true parameter values.}
\label{fig:inference}
\end{figure}

HIGlow can also be used in order to infer the underlying cosmological parameters, given a specific HI map. Bayes' law can be approximated in order to find the posterior distribution $p(z|x)$:
\begin{equation}
    p(z|x)=\frac{p(z)p(x|z)}{\intop p(x,z)dz}\approx \frac{p(z)p(x|z)}{\sum_{i=1}^N p(x|\tilde{z}_i)p(\tilde{z}_{i})}
\end{equation}
where $\tilde{z}_1,\cdots,\tilde{z}_N$ are $N$ samples from $p(z)$.

In CAMELS, the distribution over the cosmological parameters is uniform, so this is the setting we use in the following experiments, as it also follows for the test data. If a different prior over cosmological parameters is given, it can also easily be incorporated in this step.
Figure~\ref{fig:inference}-A shows examples of the posterior distributions of HI maps the model didn't see during training time. The posterior distribution in these plots is concentrated around the true parameter values and admits a measure of uncertainty, which may be useful in real experimental settings.

The posterior distribution can also be utilized in order to estimate the cosmological parameters given a new, unseen, HI map. To do so, we approximate the minimum mean-squared error (MMSE) estimator, given by $\hat{z}=\mathbb{E}_z\left[z|x\right]$, where $\hat{z}$ is the estimated value. Assuming that HIGlow correctly captures the conditional distribution, the MMSE is assured to give the most accurate estimate (in expectation). Results for parameter inference can be seen in Figure~\ref{fig:inference}; as can be seen, the predictions are typically close to the true values. This estimator achieves an RMSE of 0.058 for $\Omega_m$ and 0.065 for $\sigma_8$.

\section{Discussion}
In this work, we have introduced a new generative model, HIGlow, capable of generating high-fidelity HI maps and allows for exact likelihood estimation of high-dimensional data points. HIGlow also allows for the generation of new data points conditional on cosmological parameters, with  a similar power spectrum distribution to that of the CAMELS simulation suite. By utilizing the ability of this model to calculate the exact likelihood of maps conditional on a set of parameters, the posterior of the cosmological parameters conditioned on HI maps is made possible. This further allows for accurate parameter inference, which may prove helpful in upcoming large-scale cosmological surveys.

As HIGlow is an accurate model of the CAMELS simulation, it is our hope that HIGlow can be used to generate novel samples of HI maps under different cosmological parameters. Future work will focus on using HIGlow to forecasting for specific experiments such as CHIME~\citep{CHIME}, HIRAX~\citep{HIRAX} and SKA~\citep{SKA}. This will include a full treatment to all instrumental effects, such as angular resolution, thermal noise and foreground cleaning.

\section{Broader Impact} \label{sec:impacts}
In this work, a new generative model for HI maps in the early universe was introduced, allowing for faster generation of new HI maps as well as parameter inference. Hopefully, the methods introduced in this work will facilitate cosmological research. While in general machine learning can be misused, but we recognize no situation where the methodologies, models or implementations of this work can lead towards any negative societal impacts. 

\begin{ack}
The authors would like to acknowledge support provided by the Simons Foundation as well as  support for Program number HST-HF2-51507 provided by NASA through a grant from the Space Telescope Science Institute, which is operated by the Association of Universities for Research in Astronomy, incorporated, under NASA contract NAS5-26555.

\end{ack}


{
\small
\bibliography{biblio}

\begin{thebibliography}{21}
\providecommand{\natexlab}[1]{#1}
\providecommand{\url}[1]{\texttt{#1}}
\expandafter\ifx\csname urlstyle\endcsname\relax
  \providecommand{\doi}[1]{doi: #1}\else
  \providecommand{\doi}{doi: \begingroup \urlstyle{rm}\Url}\fi

\bibitem[{Mellema} et~al.(2013){Mellema}, {Koopmans}, {Abdalla}, {Bernardi},
  {Ciardi}, {Daiboo}, {de Bruyn}, {Datta}, {Falcke}, {Ferrara}, {Iliev},
  {Iocco}, {Jeli{\'c}}, {Jensen}, {Joseph}, {Labroupoulos}, {Meiksin},
  {Mesinger}, {Offringa}, {Pandey}, {Pritchard}, {Santos}, {Schwarz},
  {Semelin}, {Vedantham}, {Yatawatta}, and {Zaroubi}]{SKA}
Garrelt {Mellema}, L{\'e}on V.~E. {Koopmans}, Filipe~A. {Abdalla}, Gianni
  {Bernardi}, Benedetta {Ciardi}, Soobash {Daiboo}, A.~G. {de Bruyn}, Kanan~K.
  {Datta}, Heino {Falcke}, Andrea {Ferrara}, Ilian~T. {Iliev}, Fabio {Iocco},
  Vibor {Jeli{\'c}}, Hannes {Jensen}, Ronniy {Joseph}, Panos {Labroupoulos},
  Avery {Meiksin}, Andrei {Mesinger}, Andr{\'e}~R. {Offringa}, V.~N. {Pandey},
  Jonathan~R. {Pritchard}, Mario~G. {Santos}, Dominik~J. {Schwarz}, Benoit
  {Semelin}, Harish {Vedantham}, Sarod {Yatawatta}, and Saleem {Zaroubi}.
\newblock {Reionization and the Cosmic Dawn with the Square Kilometre Array}.
\newblock \emph{Experimental Astronomy}, 36\penalty0 (1-2):\penalty0 235--318,
  August 2013.
\newblock \doi{10.1007/s10686-013-9334-5}.

\bibitem[DeBoer et~al.(2017)DeBoer, Parsons, Aguirre, Alexander, Ali,
  Beardsley, Bernardi, Bowman, Bradley, Carilli, et~al.]{HERA}
David~R DeBoer, Aaron~R Parsons, James~E Aguirre, Paul Alexander, Zaki~S Ali,
  Adam~P Beardsley, Gianni Bernardi, Judd~D Bowman, Richard~F Bradley, Chris~L
  Carilli, et~al.
\newblock Hydrogen epoch of reionization array (hera).
\newblock \emph{Publications of the Astronomical Society of the Pacific},
  129\penalty0 (974):\penalty0 045001, 2017.

\bibitem[van Haarlem et~al.(2013)van Haarlem, Wise, Gunst, Heald, McKean,
  Hessels, de~Bruyn, Nijboer, Swinbank, Fallows, et~al.]{LOFAR}
Michael~P van Haarlem, Michael~W Wise, AW~Gunst, George Heald, John~P McKean,
  Jason~WT Hessels, A~Ger de~Bruyn, Ronald Nijboer, John Swinbank, Richard
  Fallows, et~al.
\newblock Lofar: The low-frequency array.
\newblock \emph{Astronomy \& astrophysics}, 556:\penalty0 A2, 2013.

\bibitem[Ivezi{\'c} et~al.(2019)Ivezi{\'c}, Kahn, Tyson, Abel, Acosta, Allsman,
  Alonso, AlSayyad, Anderson, Andrew, et~al.]{LSST}
{\v{Z}}eljko Ivezi{\'c}, Steven~M Kahn, J~Anthony Tyson, Bob Abel, Emily
  Acosta, Robyn Allsman, David Alonso, Yusra AlSayyad, Scott~F Anderson, John
  Andrew, et~al.
\newblock Lsst: from science drivers to reference design and anticipated data
  products.
\newblock \emph{The Astrophysical Journal}, 873\penalty0 (2):\penalty0 111,
  2019.

\bibitem[Spergel et~al.(2015)Spergel, Gehrels, Baltay, Bennett, Breckinridge,
  Donahue, Dressler, Gaudi, Greene, Guyon, et~al.]{Roman}
D~Spergel, N~Gehrels, C~Baltay, D~Bennett, J~Breckinridge, M~Donahue,
  A~Dressler, BS~Gaudi, T~Greene, O~Guyon, et~al.
\newblock Wide-field infrarred survey telescope-astrophysics focused telescope
  assets wfirst-afta 2015 report.
\newblock \emph{arXiv preprint arXiv:1503.03757}, 2015.

\bibitem[Dor{\'e} et~al.(2014)Dor{\'e}, Bock, Ashby, Capak, Cooray, de~Putter,
  Eifler, Flagey, Gong, Habib, et~al.]{SPHEREx}
Olivier Dor{\'e}, Jamie Bock, Matthew Ashby, Peter Capak, Asantha Cooray,
  Roland de~Putter, Tim Eifler, Nicolas Flagey, Yan Gong, Salman Habib, et~al.
\newblock Cosmology with the spherex all-sky spectral survey.
\newblock \emph{arXiv preprint arXiv:1412.4872}, 2014.

\bibitem[Racca et~al.(2016)Racca, Laureijs, Stagnaro, Salvignol, Alvarez,
  Criado, Venancio, Short, Strada, B{\"o}nke, et~al.]{Euclid}
Giuseppe~D Racca, Ren{\'e} Laureijs, Luca Stagnaro, Jean-Christophe Salvignol,
  Jos{\'e}~Lorenzo Alvarez, Gonzalo~Saavedra Criado, Luis~Gaspar Venancio, Alex
  Short, Paolo Strada, Tobias B{\"o}nke, et~al.
\newblock The euclid mission design.
\newblock In \emph{Space telescopes and instrumentation 2016: optical,
  infrared, and millimeter wave}, volume 9904, pages 235--257. SPIE, 2016.

\bibitem[Villaescusa-Navarro et~al.(2021{\natexlab{a}})Villaescusa-Navarro,
  Angl{\'e}s-Alc{\'a}zar, Genel, Spergel, Li, Wandelt, Nicola, Thiele, Hassan,
  Matilla, et~al.]{CNN1}
Francisco Villaescusa-Navarro, Daniel Angl{\'e}s-Alc{\'a}zar, Shy Genel,
  David~N Spergel, Yin Li, Benjamin Wandelt, Andrina Nicola, Leander Thiele,
  Sultan Hassan, Jose Manuel~Zorrilla Matilla, et~al.
\newblock Multifield cosmology with artificial intelligence.
\newblock \emph{arXiv preprint arXiv:2109.09747}, 2021{\natexlab{a}}.

\bibitem[Hassan et~al.(2020)Hassan, Andrianomena, and Doughty]{CNN2}
Sultan Hassan, Sambatra Andrianomena, and Caitlin Doughty.
\newblock Constraining the astrophysics and cosmology from 21 cm tomography
  using deep learning with the ska.
\newblock \emph{Monthly Notices of the Royal Astronomical Society},
  494\penalty0 (4):\penalty0 5761--5774, 2020.

\bibitem[Goodfellow et~al.(2020)Goodfellow, Pouget-Abadie, Mirza, Xu,
  Warde-Farley, Ozair, Courville, and Bengio]{GAN}
Ian Goodfellow, Jean Pouget-Abadie, Mehdi Mirza, Bing Xu, David Warde-Farley,
  Sherjil Ozair, Aaron Courville, and Yoshua Bengio.
\newblock Generative adversarial networks.
\newblock \emph{Communications of the ACM}, 63\penalty0 (11):\penalty0
  139--144, 2020.

\bibitem[Kingma and Welling(2013)]{VAE}
Diederik~P Kingma and Max Welling.
\newblock Auto-encoding variational bayes.
\newblock \emph{arXiv preprint arXiv:1312.6114}, 2013.

\bibitem[Dinh et~al.(2014)Dinh, Krueger, and Bengio]{dinh2014nice}
Laurent Dinh, David Krueger, and Yoshua Bengio.
\newblock Nice: Non-linear independent components estimation.
\newblock \emph{arXiv preprint arXiv:1410.8516}, 2014.

\bibitem[Kingma and Dhariwal(2018)]{Glow}
Durk~P Kingma and Prafulla Dhariwal.
\newblock Glow: Generative flow with invertible 1x1 convolutions.
\newblock \emph{Advances in Neural Information Processing Systems}, 31, 2018.

\bibitem[Lu and Huang(2020)]{cGlow}
You Lu and Bert Huang.
\newblock Structured output learning with conditional generative flows.
\newblock In \emph{Proceedings of the AAAI Conference on Artificial
  Intelligence}, volume~34, pages 5005--5012, 2020.

\bibitem[Lugmayr et~al.(2020)Lugmayr, Danelljan, Gool, and Timofte]{SRFlow}
Andreas Lugmayr, Martin Danelljan, Luc~Van Gool, and Radu Timofte.
\newblock Srflow: Learning the super-resolution space with normalizing flow.
\newblock In \emph{European Conference on Computer Vision}, pages 715--732.
  Springer, 2020.

\bibitem[Papamakarios et~al.(2021)Papamakarios, Nalisnick, Rezende, Mohamed,
  and Lakshminarayanan]{papamakarios2021normalizing}
George Papamakarios, Eric~T Nalisnick, Danilo~Jimenez Rezende, Shakir Mohamed,
  and Balaji Lakshminarayanan.
\newblock Normalizing flows for probabilistic modeling and inference.
\newblock \emph{J. Mach. Learn. Res.}, 22\penalty0 (57):\penalty0 1--64, 2021.

\bibitem[Hassan et~al.(2021)Hassan, Villaescusa-Navarro, Wandelt, Spergel,
  Angl{\'e}s-Alc{\'a}zar, Genel, Cranmer, Bryan, Dav{\'e}, Somerville,
  et~al.]{HIFlow}
Sultan Hassan, Francisco Villaescusa-Navarro, Benjamin Wandelt, David~N
  Spergel, Daniel Angl{\'e}s-Alc{\'a}zar, Shy Genel, Miles Cranmer, Greg~L
  Bryan, Romeel Dav{\'e}, Rachel~S Somerville, et~al.
\newblock Hiflow: Generating diverse hi maps conditioned on cosmology using
  normalizing flow.
\newblock \emph{arXiv preprint arXiv:2110.02983}, 2021.

\bibitem[Villaescusa-Navarro et~al.(2021{\natexlab{b}})Villaescusa-Navarro,
  Angl{\'e}s-Alc{\'a}zar, Genel, Spergel, Somerville, Dave, Pillepich,
  Hernquist, Nelson, Torrey, et~al.]{CAMELS}
Francisco Villaescusa-Navarro, Daniel Angl{\'e}s-Alc{\'a}zar, Shy Genel,
  David~N Spergel, Rachel~S Somerville, Romeel Dave, Annalisa Pillepich, Lars
  Hernquist, Dylan Nelson, Paul Torrey, et~al.
\newblock The camels project: Cosmology and astrophysics with machine-learning
  simulations.
\newblock \emph{The Astrophysical Journal}, 915\penalty0 (1):\penalty0 71,
  2021{\natexlab{b}}.

\bibitem[Dinh et~al.(2016)Dinh, Sohl-Dickstein, and Bengio]{RealNVP}
Laurent Dinh, Jascha Sohl-Dickstein, and Samy Bengio.
\newblock Density estimation using real nvp.
\newblock \emph{arXiv preprint arXiv:1605.08803}, 2016.

\bibitem[Amiri et~al.(2022)Amiri, Bandura, Boskovic, Chen, Cliche, Deng,
  Denman, Dobbs, Fandino, Foreman, et~al.]{CHIME}
Mandana Amiri, Kevin Bandura, Anja Boskovic, Tianyue Chen, Jean-Fran{\c{c}}ois
  Cliche, Meiling Deng, Nolan Denman, Matt Dobbs, Mateus Fandino, Simon
  Foreman, et~al.
\newblock An overview of chime, the canadian hydrogen intensity mapping
  experiment.
\newblock \emph{The Astrophysical Journal Supplement Series}, 261\penalty0
  (2):\penalty0 29, 2022.

\bibitem[Newburgh et~al.(2016)Newburgh, Bandura, Bucher, Chang, Chiang, Cliche,
  Dav{\'e}, Dobbs, Clarkson, Ganga, et~al.]{HIRAX}
LB~Newburgh, K~Bandura, MA~Bucher, T-C Chang, HC~Chiang, JF~Cliche, R~Dav{\'e},
  M~Dobbs, C~Clarkson, KM~Ganga, et~al.
\newblock Hirax: a probe of dark energy and radio transients.
\newblock In \emph{Ground-based and Airborne Telescopes VI}, volume 9906, pages
  2039--2049. SPIE, 2016.

\end{thebibliography}
}

\appendix



\end{document}